# MRGEN: A CONCEPTUAL FRAMEWORK FOR LLM-POWERED MIXED REALITY AUTHORING TOOLS FOR EDUCATION

Mohammed Oussama Seddini
*LIUM, Le Mans Université*
*72085 Le Mans, France*

Ngoc Luyen Le
*Gamaizer*
*93340 Le Raincy, France*

Mohamed Ez-Zaouia
*IRISA, Université de Rennes*
*35000 Rennes, France*

Iza Marfisi-Schottman
*LIUM, Le Mans Université*
*72085 Le Mans, France*

**ABSTRACT**

Mixed Reality (MR) offers immersive and multimodal opportunities for education but remains difficult for teachers to author without technical expertise. We propose MRGEN, a conceptual framework for LLM-powered authoring tools to support teachers in creating MR learning activities that work on mobile devices (tablets and smartphones). MRGEN articulates three axes: Learning Objectives, MR Modality, and GAI Assistance. To validate our framework, we implemented a prototype based on the open-source MIXAP authoring platform and conducted a user study with 24 participants. Results show that LLM-powered authoring reduced task duration by 36% on average, and that over 90% of participants found the AI support helpful for brainstorming, structuring, and aligning content with their learning goals. These findings yielded very promising results for future AI-assisted MR authoring tools.

## 1. INTRODUCTION

Mixed Reality (MR) offers new opportunities for education by enabling immersive, multimodal learning experiences with widespread tablets and smartphones (Zammit, 2025). Research has shown that MR can increase learner engagement and improve understanding of complex concepts by combining visual, audio, and kinesthetic modalities (Huang & Tseng, 2025). However, creating MR activities remains difficult for many teachers, who often lack programming skills or the time required to design effective experiences (Nguyen et al., 2025). Authoring tools have emerged to reduce these technical barriers (Liu, 2024). Yet, teachers still face challenges (Ez-Zaouia, Marfisi-Schottman & Mercier, 2023). They struggle to (1) find inspiration for creating activities suitable for their classes, and (2) create the multimodal resources, such as images, 3D objects, and audio recordings, and (3) align MR content with learning objectives.

Recent advancements in Generative Artificial Intelligence (GAI) or large language models (LLMs) present a promising solution (Hemminki-Reijonen et al., 2025). By enabling the creation of content from natural language prompts, GAI can support teachers in designing MR activities. However, despite growing interest in GAI for education, its integration into MR authoring tools remains underexplored. Moreover, there is limited guidance on how GAI assistance should be structured to meet immersive learning goals.

To address this gap, we examine, in this paper, how AI can simplify the process for teachers to create MR activities, specifically: which learning objectives, MR modalities, and AI assistance to facilitate the creation of MR educational activities? (RQ1) and are these features perceived as useful by teachers? (RQ2).

Section 2 presents prior related work on MR and GAI for education. Section 3 presents the MRGEN framework, articulating three intersecting design axes: Learning Objectives, MR Modality and GAI Assistance. The practical feasibility of this framework is also demonstrated through a prototype implementation (section

4). In Section 5, the user study with 24 participants is presented, including a few results. Finally, section 5 presents the conclusion and perspectives of this research.

## 2. BACKGROUND AND RELATED WORK

MR also enables learners to visualize abstract phenomena and interact with digital content in real-world contexts, supporting deeper learning in fields such as STEM, language education, and vocational training (Radianti et al., 2020). MR can provide access to resources that would be physically unavailable to learners, such as exploring the solar system or visualizing magnetic fields. However, authoring MR activities is complex for teachers due to the programming skills required. Authoring Tools offer a promising solution by providing simple activity editors for non-programmers like teachers. However, a review of 21 MR authoring tools found that none fully met teachers' needs (Ez-Zaouia et al., 2022). Teachers still struggle to (1) find inspiration for creating activities suitable for their classes, and (2) create the multimodal resources, such as images, 3D objects, and audio recordings, and (3) align MR content with learning objectives.

GAI or LLMs (e.g., GPT-4, DALL·E, Whisper), can support teachers in creating multimedia content, such as text, images, audio, video, and personalized feedback from natural language prompts (Almeman et al., 2025) GAI also acts as a creative partner, suggesting activities aligned with learning goals (Yannier et al., 2024). However, these benefits come with challenges. Ensuring pedagogical alignment, avoiding bias, and safeguarding data privacy are essential. Teachers also need support in prompt engineering and critical evaluation of AI outputs. Emerging tools like AirSketch (Lim et al., 2024) or Unity Muse (Unity Technologies, 2025) begin to integrate GAI within VR/MR environments. However, they often target technical creators and remain limited in terms of pedagogical structuring or alignment with learning objectives.

In this paper, we aim to combine both GAI assistance (LLMs) and visual authoring techniques to support teachers in creating MR learning activities. We build upon MIXAP, an open-source authoring tool, introduced recently to facilitate the creation of educational activities involving MR (Ez-Zaouia, Marfisi-Schottman & Mercier, 2023; MIXAP Team, 2025). Developed with feedback from about twenty pilot teachers who tested it in various classes (from kindergarten to high school), MIXAP offers several predefined MR learning activity models. It also integrates a four-step process for creating MR activities with visual interactions similar to those found in PowerPoint. Considering the validated approach of this tool and its open-source nature, we choose to examine our GAI assistance approach by adding AI modules to MIXAP. These modules extend the platform with LLM-powered authoring features that help teachers generate, structure, and adapt MR content through natural language prompts.

## 3. CONCEPTUAL FRAMEWORK AND IMPLEMENTATION

To support teachers in designing pedagogical MR learning activities using GAI (RQ1), we propose MRGEN, a conceptual framework with three axes: (1) Learning Objective, (2) MR Modality, and (3) AI Assistance (Figure 1). We formulated our framework based on prior research as well as an open dataset that describes 18 MR activities, imagined by 18 teachers (Liu, 2024, Ez-Zaouia et al., 2022, Marfisi-Schottman 2022).

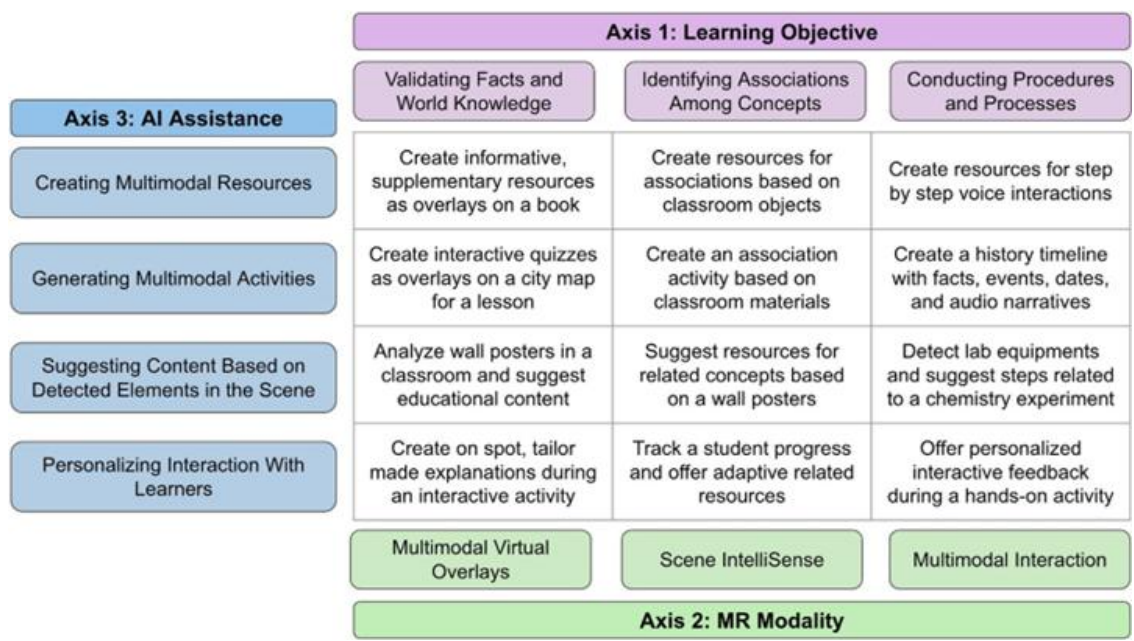

Figure 1: The three axes of the MRGEN conceptual framework with a few examples.

## 3.1 AXIS 1 – LEARNING OBJECTIVE (LO)

Drawing on Bloom's Taxonomy (Bloom et al., 1956) and teacher's intents for MR [11], we identify
three recurrent categories of learning objectives that frequently guide MR activity design.

- **LO1 - Validating facts or knowledge.** Bloom's "understand" and "evaluate": Activities that help learners verify their knowledge and assess their understanding through MR interactions (e.g., "placing historical events on a timeline or a map").

- **LO2 - Identifying associations among concepts.** Bloom's "understand" and "analyze": Activities that help learners break down information into parts to explore relationships and connect concepts (e.g., *"associating words to images or audios"*).

- **LO3 - Conducting procedures and processes.** Bloom's "apply", "analyze", and "evaluate": Activities that guide learners to use knowledge and concepts in new situations via sequential tasks (e.g., *"following a step-by-step procedure"*).

## 3.2 AXIS 2 – MR MODALITY (MM)

Drawing on TPACK framework, that emphasizes the alignment between technology, pedagogy, and content (Mishra & Koehler, 2006), Mayer's multimedia learning (ML), that highlights the importance of using of multiple sensory channels (Mayer, 2009), and teachers MR modalities (Mishra & Koehler, 2006), we identify three recurring types of modalities.

- **MM1 - Multimodal Virtual Overlays.** Overlaying digital content (text, images, audio.) on physical materials, such as books, posters or worksheets (e.g., "linking vocabulary to storybook images or adding legends to a geographic map").

- **MM2 - Scene IntelliSense.** Using MR systems to interpret the physical environment through object or scene recognition and changing the content based on these real-world inputs (e.g., "scanning a school mural to display animal names and their sounds, or augmenting plants in a schoolyard").

- **MM3 - Multimodal Interaction.** Interacting with virtual elements using touch, gestures, voice, haptic feedback, zoom, or rotation, invites learners to engage with the content (e.g., "rotating 3D tectonic plates while listening to explanations, or using voice commands to trigger MR augmentations").

## 3.3 AXIS 3 – AI ASSISTANCE (AA)

Drawing on the SAMR model (Substitution, Augmentation, Modification, Redefinition) (Hamilton, Rosenberg & Akcaoglu, 2016), we identify four levels of GAI assistance:

- **AA1 - Creating Multimodal Resources.** Using GAI to create license-free resources, such as images or converting text to speech can help teachers with these repetitive and time-consuming tasks (e.g., "generating audio instructions for learners with reading difficulties").

- **AA2 - Generating Multimodal Activities.** Using GAI to generate full MR activity based on instructional prompts, educational content, and learning goals (e.g., "detecting objects in a book and automatically generating labels, audio files, and images for a vocabulary learning task").

- **AA3 - Context-Aware Design Support.** Using GAI to generate real-time contextual information such as camera input or spatial recognition to suggest relevant content or activities (e.g., "recognizing printed materials on a table and proposing a map-based timeline activity for a history lesson").

- **Personalizing Interaction with Learners.** Using GAI to offer personalized MR activities in response to individual learner needs, behaviors, or performances (e.g., "allowing students to interact with portraits of the British royalty by asking questions and receiving tailored responses as text or audio in English").

## 4. MRGEN PROTOTYPE

To provide a technical validation of our conceptual model, we implemented three interconnected GAI modules in MIXAP:

*AI-Prompter:* enables creation of multimodal resources, i.e., images, audio, and text using text prompts (AA1). As Figure 2-(a) shows, after taking a photo of a lion (image marker), teachers might want to add additional information as MR content. They can select "add Image" menu (1), select AI (2), enter a prompt (e.g., "a photo of a lion") (3), and the generated content will appear as an MR resource (4). The generated MR resources can be (5) modified, resized, and arranged by the teacher on the canvas.

*AI-Chat:* provides suggestions for creating MR activities, based on the MR marker image (AA1 and AA2). As Figure 2-(b) shows, a dialogue box appears (1), it automatically analyzes educational resources provided by the teacher (e.g., world map) and proposes activity ideas (3), namely describing the image, generating historical facts related to different countries (4). Similarly, teachers can modify, resize, and reposition the generated MR resources.

***AI-Tutor:*** offers personalized learning experiences by responding and adapting to learner inputs and providing contextual information (AA3 and AA4). This module becomes active when the learner runs the activity. As Figure 2-(c) shows, learners can interact with AI-Tutor using their fingers and question cards. For example, a learner can ask questions about one of the continents on the map by pointing at the area of interest with their finger and using corresponding cards (1). The AI processes the scene and provides contextual responses (2). Cards can be provided by the teacher or written by learners.

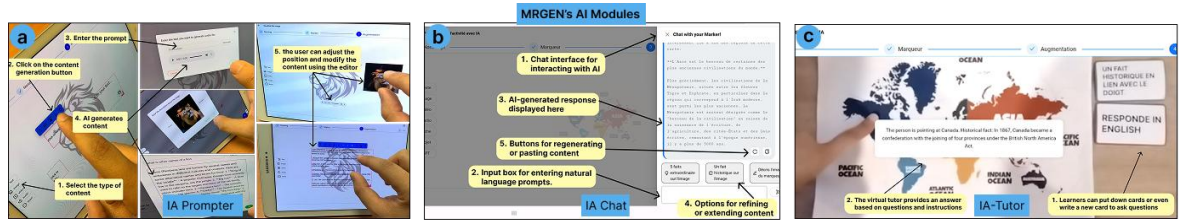

Figure 2: Overview of participants' use of the MRGEN framework and examples of AI-assisted productions. (a) AI Prompter: Teachers use text prompts to generate multimodal MR resources (images, audio, text) that are directly added to the scene. (b) AI Chat: Teachers receive contextual suggestions based on the image marker and can refine content through prompt iterations. (c) AI Tutor: Learners interact with MR content using physical cues (e.g., pointing, cards), triggering personalized responses from the AI. (d) User Production Matrix: Summary of participant activities—Activity 1 (manual authoring) and Activity 2 (with GAI)—with dots representing submitted outputs. Themes range from world geography to biology and social studies. (e) Examples of AI-generated content: For each participant, we show the prompt, selected outputs (image/text/audio), and representative content generated by GAI. These examples highlight the diversity and richness of multimodal outputs produced through the MRGEN framework.

## 5. USER STUDY

For empirical validation, we ran a user study, focusing on the usability and usefulness of our approach.

### 5.1 Method

***Participants.*** The study involved 24 participants, including 16 in-service teachers recruited from primary, secondary, and vocational education and 8 university students.

***Procedure.*** Participants were asked to create three activities using MIXAP. They had to create the first activity without any GAI support (which served as a baseline). They were required to create a MR activity on a world map with at least one text, one image, and one audio element to support a pedagogical objective. For creating the second activity, they used AI-Prompter and AI-Chat to generate textual, visual, and audio content. To create the third activity, the participants were asked to use AI-Tutor with the world map.

***Data collection.*** Each session lasted approximately 60 minutes. We collected 18 responses to a pre-survey where the participants described the learning objective and initial idea for their activity and 24 responses to a post-survey designed to assess usability, perceived usefulness of the GAI tools, and overall qualitative feedback. Moreover, we collected the interaction logs and the 72 MR activities produced during the study (Figure 2-(d) presents illustrative examples of these).

***Data Analysis***. Interaction logs were analyzed using non-parametric statistical methods. Specifically, Kruskal–Wallis tests (Kruskal & Wallis, 1952) were applied to identify significant differences in AI modules, followed by Wilcoxon pairwise comparisons with Holm correction and effect size calculations. Qualitative data, including survey responses and user-generated content, were analyzed through thematic analysis. The first author conducted multiple coding iterations, which were then discussed with the co-authors.

### 5.2 Results

This comparative study revealed several interesting results. Because of the limited space in this paper, we will only present three of our findings:

***Non-GAI VS. GAI-Supported Authoring.*** We compared the activities completed without GAI assistance, and the ones completed with AI-Prompter and AI-Chat (AI-Tutor was excluded from this analysis, as it does not involve content creation). On average, activities completed with GAI were designed more quickly (M = 12.2 min, SD = 4.8) than those completed without (M = 19.1 min, SD = 5.1). This can be attributed to the automation of resource generation and idea suggestion, which enabled participants to focus more on instructional design rather than on manually searching for or creating content. In total, participants submitted 142 prompts, which resulted in 140 generated outputs, including 47 regenerations with slightly different prompts. This suggests that, for some users, refining (editing) and customizing GAI-generated content required multiple cycles. This was also reflected in the interaction logs: 6 out of 24 participants spent more time on the GAI-supported activity, with an average 6.7 minutes more compared to the non-GAI activities. This extended duration was likely due to deeper engagement with the design process. Rather than using GAI solely to reduce effort, these participants appeared to leverage its capabilities to enrich the complexity and multimodality of their activities. Indeed, as illustrated in Table D in Figure 2, they explored a broader range of formats (e.g., text, images, audio) whereas activities completed without GAI were generally simpler.

***Learning Objectives and Outcomes.*** Participants designed learning activities in a large variety of domains (see Figure 2-(e)). The most represented domains were Geography (n = 13), with participants creating activities on population distribution, natural resource mapping, and comparative cartographic analysis (e.g., P09, P10, P21). History and Biology were each addressed by four participants, focusing respectively on historical empires and migration patterns (e.g., P12, P24) and on fauna classification across continents (e.g., P02, P07). The remaining themes were covered by individual participants: AI Technology (P22), Human Anatomy (P01), and

Social Studies (P14). These results highlight the adaptability of GAI in supporting diverse pedagogical intentions.

***Users' Perspectives.*** Post-survey responses indicate that participants generally perceived the GAI tools as beneficial. Over 90% of respondents agreed or strongly agreed that AI enhanced the pedagogical quality of their activities, supported content generation, and saved time during the design process. P15 noted: "*It was intuitive and efficient; I could see my generated content come together in just a few minutes* —P15." Specifically, 92% of participants reported that the GAI generated relevant content, while an equivalent proportion acknowledged gains in time efficiency and support in structuring educational materials. Participant P12 commented: "*The AI helped me quickly generate text and audio content* —P12." AI-Tutor was also perceived as particularly effective. As P03 expressed: "*The handwritten card recognition is a great feature* —P03." Finally, suggestions for improvement included the addition of 3D content generation and enhanced customization of prompts.

## 6. CONCLUSION

Regarding RQ1, we introduced MRGEN, a conceptual framework integrating GAI into MR authoring tools to assist teachers in creating multimodal and pedagogically aligned learning activities. Our framework articulates three axes: Learning Objectives, MR Modality, and GAI Assistance. Regarding RQ2, we carried out a two-fold approach to validate our framework. We implemented a first prototype for a technical validation, where we built upon previous research, in particular the MIXAP authoring tool. We also carried out empirical validation via a user study with 24 participants.

The results were promising. GAI-assisted authoring led to a 36% reduction in task duration on average, and over 90% of participants found the AI helpful for brainstorming, structuring, and aligning content with their learning goals. Many also used GAI to enrich their activities with more modalities (text, images, audio) generating diverse productions, confirming its value beyond simple efficiency gains (Hamilton, Rosenberg & Akcaoglu, 2016; Nguyen et al., 2025). In light of previous results on MR authoring for classroom use (e.g., Mercier et al., 2023, 2024), our approach can reduce the time and effort required to create multimodal resources for MR learning activities, allowing teachers with less technical expertise to focus more closely on pedagogical instruction.

Some limitations were observed. The quality of GAI-generated images was occasionally affected by vague prompts, underscoring the need for prompt refinement tools. Moreover, a few participants misunderstood the audio capabilities, expecting generative sound synthesis when only text-to-speech was supported. Questions also emerged around how to better align GAI outputs with learning objectives. These findings highlight an initial validation of the GAI's utility while underlining the need for expanded, adaptive, and pedagogically grounded functionalities in future developments.

## ACKNOWLEDGEMENT

We warmly thank the consortium Ikigai, led by the association Games for Citizens, the company Gamaizer, as well as the FORTEIM project (laureate of the AMI CMA France 2030), for their support. This work was financed by the Mixed Reality Authoring Applications for and by Teachers in Europe (MIXAP-EU) project (2024-1-FR01-KA220-SCH-000245309) and is co-funded by the European Union under the Erasmus+ Program, Cooperation Partnerships in School Education (KA220-SCH).

## REFERENCES

Almeman, K. et al. (2025). The integration of AI and Metaverse in education: A systematic literature review. Applied Sciences, 15(2), 863.

Bloom, B. S. et al. (1956). Taxonomy of educational objectives: The classification of educational goals. Handbook I: Cognitive domain.

Ez-Zaouia, M. et al. (2022). A design space of educational authoring tools for augmented reality. In International Conference on Games and Learning Alliance, pp. 258–268.

Ez-Zaouia, M., Marfisi-Schottman, I. & Mercier, C. (2023). Authoring tools: The road to democratizing augmented reality for education. In Proceedings of the 15th International Conference on Computer Supported Education, Vol. 1, pp. 115–127.

Hamilton, E., Rosenberg, J. & Akcaoglu, M. (2016). The Substitution Augmentation Modification Redefinition (SAMR) model: A critical review and suggestions for its use. TechTrends, 60.

Hemminki-Reijonen, U. et al. (2025). Design of generative AI-powered pedagogy for virtual reality environments in higher education. npj Science of Learning, 10(1).

Huang, T. & Tseng, H.-C. (2025). Extended reality in applied sciences education: A systematic review. Applied Sciences, 15(7), 4038.

Kruskal, W. H. & Wallis, W. A. (1952). Use of ranks in one-criterion variance analysis. Journal of the American Statistical Association, 47(260), 583–621.

Lim, H. X. G. et al. (2024). AirSketch: Generative motion to sketch. arXiv preprint, arXiv:2407.08906.

Liu, X. (2024). Use of a mixed-reality creative environment in design education based on Microsoft HoloLens 2. Procedia Computer Science, 240.

Marfisi-Schottman, I. (2022). Teachers paper-prototyping augmented reality activities for classroom use. Mendeley Data, V1.

Mayer, R. E. (2009). Multimedia learning (2nd ed.).

Mercier C, Marfisi-Schottman I, Ez-Zaouia M, Deshayes D. La réalité augmentée en classe au service des apprentissages des élèves. Médiations et médiatisations-Revue internationale sur le numérique en éducation et communication. 2023 Jun 30(15):78-98.

Mercier, C., Marfisi, I. and Ez-Zaouia, M., 2024. La réalité augmentée au service de la médiation-remédiation cognitive. Spirale-Revue de recherches en éducation, 73(1), pp.203-213.

Mishra, P. & Koehler, M. J. (2006). Technological Pedagogical Content Knowledge: A framework for teacher knowledge. Teachers College Record, 108(6), 1017–1054.

MIXAP Team (2025). MIXAP: Open-source Mixed Reality authoring platform. Available at: https://forge.apps.education.fr/lium/mixap
 (Accessed July 2025).

Nguyen, A. et al. (2025). Designing embodied generative artificial intelligence in mixed reality for active learning in higher education. Innovations in Education and Teaching International, 1.

Radianti, J. et al. (2020). A systematic review of immersive virtual reality applications for higher education: Design elements, lessons learned, and research agenda. Computers & Education, 147, 103778.

Unity Technologies (2025). Unity Muse. Available at: https://unity.com/fr/products/muse
 (Accessed July 2025).

Yannier, N. et al. (2024). AI adaptivity in a mixed-reality system improves learning. International Journal of Artificial Intelligence in Education, 34(4), 1541.

Zammit, J. P. (2025). Education 4.0 for Industry 4.0: A mixed reality framework for workforce readiness in manufacturing. Multimodal Technologies and Interaction, 9(5), 43.